\documentstyle[art11]{article}
\newcommand{\EqRef}[1]{ Eq. (\ref{#1})}
\newcommand{\EqLabel}[1]{\label{#1}}
\begin{document}
\pagestyle{plain}
\begin{center}
{\LARGE
 Hamiltonian path integral quantization in polar coordinates} \\
\vspace{1in}
A.K.Kapoor\\ School of Physics\\ University of Hyderabad \\
 Hyderabad 500046 INDIA\\
\vspace{0.5cm}
and\\
\vspace{0.5cm}
Pankaj Sharan \\
Physics Department\\
Jamia Milia Islamia, Jamia Nagar 
\\New Delhi 110025, INDIA
\\
\vspace{1in}
\underline {ABSTRACT}\\
\end{center}

Using  a  scheme  proposed  earlier  we  set up Hamiltonian path integral 
quantization   for   a   particle   in  two  dimensions  in  plane  polar  
coordinates.This  scheme  uses  the  classical  Hamiltonian,  without any  
$O(\hbar^2)$  terms, in the polar varivables. We show that the propagator 
satisfies the correct Schr\"{o}dinger equation.

\newpage
\section{Introduction}
The Feynman path integral scheme gives an important route to quantization 
\cite{REVART}.  That  in  non-cartesian  coordinates  one  needs  to  add  
$O(\hbar^2)$  terms  to the potential to arrive at correct path integral, 
was  at  first  demonstrated  in polar coordinates by Edwards and Gulyaev 
\cite{EDWARD}   who   also  computed  the  free  particle  propagator  in  
$r,\theta$  variables using the path integrals. However, for systems with 
finite  degrees  of freedom and with Lagrangians quadratic in velocities, 
the   scheme   of   Pauli   and   DeWitt  \cite{BSDEWITT,PAULI}  has  the  
distinguishing  feature  that  the  Lagrangian path integral quantization 
method  can  be  set  up  consistently  in  arbitrary coordinates without 
addition  of ad hoc $O(\hbar^2)$ terms to te potential. However, the same 
in  not  true for the Hamiltonian path integral quantization. It has been 
known  that $O(\hbar^2)$ terms must be added to the classical Hamiltonian 
in  order  to  arrive  at  the  correct  quantization  from  most  of the  
available the Hamiltonian path integral schemes.
\cite{PEAK,ARTHURS,GERVIAS,LANGGUTH}

A Hamiltonian path integral quantization scheme was given by one of us in 
\cite{AKK1}.  This  scheme  is  a  natural  generalization  of the Pauli-
DeWitt's   scheme   for   the   Lagrangian   formulation.   However,  the  
Hamiltonianpath  integral  suggested  in  \cite{AKK1}  failed to give the 
correct  Schr\"{o}dinger  equation  even  for  the  free  particle in two 
dimension  if  one  used  the  classical  Hamiltonian  in the plane polar 
coordinates;  this  scheme  too  required addition of ad hoc $O(\hbar^2)$ 
terms to the classical Hamiltonian. 

It  may  be  appropriate  to  recall  at  this  stage  that the canonical 
quantization  procedure  too  does  not  give  the  expected  Hamiltonian  
operator  as $-{\hbar^2\over 2m}\nabla^2 $ in the $r\theta$ coordinates . 
To  see  this  note  that  the  classical  Hamiltonian $ {P^2 \over 2m} + 
{p^2\over  2  m  r^2}$  does  not  contain a product of two non commuting 
factors.   The   canonical  quantization  gives  the  momentum  operators  
conjugate to $r , \theta$ as 

\begin{equation}
\hat{P}  =  -  i  \hbar  \frac{\partial}{\partial r} + {i \hbar\over 2r } 
,\qquad \qquad 
\hat{p} = - i \hbar \frac{\partial}{\partial \theta}
\end{equation}

Replacing  the  c-number variables by corresponding operators the quantum 
mechanical Hamiltonian as the operator is easily seen to be

\begin{eqnarray}
\hat{H} =  - \frac{\hbar^2}{2m}
\left(\frac{\partial^2 }{\partial r^2} + {1\over r} \frac {\partial 
}{\partial r} + \frac{\partial^2 }{\partial \theta^2} 
\right)+ {\hbar^2 \over 8 m r^2}
= - {\hbar^2 \over 2 m }\nabla^2 + {\hbar^2 \over 8 m r^2}
\end{eqnarray}

It  was  soon  realized  that  it  is possible to modify the formalism of 
\cite{AKK1}  by incorporating the idea of local scaling of time which had 
been found a useful technique in exact evaluation of path integratals for 
several  potential  problems  \cite{HOAKIRA,AKIRA}.  In \cite{AKK2} a new 
scheme  of  Hamiltonian  path integration was suggested incorporating the 
idea  of local scaling of time in the Hamiltonian path integral method of 
\cite{AKK1}.  This  modified  scheme of Hamiltonian path integration with 
scaling  was  further studied in \cite{AKKPS1,AKKPS2}. Working within the 
Hamiltonian  path  integral framework, an important feature of the scheme 
of \cite{AKK2} is that one can use the classical Hamiltonian in arbitrary 
coordinates  and  still  arrive  at the correct Hamiltonian path integral 
representation  for  the  quantum  mechanical  propapgator.  This  is  in  
contrast  to  the  well  known  fact  that  in  all  the  other exisiting 
Hamiltonian  path integral schemes where one is required to add $\hbar^2$ 
terms to obtain the correct Schr\"{o}dinger equation in coordinates other 
than  cartesian  coordinates;  such  terms  being absent in the cartesian 
coordinates only. 

In  this  paper we shall describe the Hamiltonian path integral scheme of 
\cite{AKK2,AKKPS1,AKKPS2}  and set up  the propagator for a free particle 
in   two   dimensions   in   plane   polar  coordinates  and  derive  the  
Schr\"{o}dinger equation for the propagator. This was after all the first 
example where the need for addition of $\hbar^2$ terms to the hamiltonian 
was  demonstrated  \cite{EDWARD}.  We also hope that this paper will make  
the  formalism,  the  methods   and  the  results  of  our earlier papers 
transparent.  In  Sec.  2  we  summaize  the  Haimltonian  path  integral  
quantization scheme of \cite{AKK2,AKKPS1} and in Sec 3 we set up the path 
integral  for  propagator  of a particle in two dimensions in plane polar 
coordinates  and  show  that  it  satisfies  the  correct Schr\"{o}dinger 
equation. 

\section{Hamiltonian path integral quantization:} 

In  this  section,  at  first,  we  shall briefly recall Lagranigian path 
integral  as  given  in  \cite{BSDEWITT,PAULI}.We  summarize the steps of 
construction  of  the  Hamiltonian  path  integral representation for the 
propagator in arbitrary coordinates from \cite{AKK2}. 
\vspace{0.5cm}

\noindent
{\bf Lagrangian path intgeral:} Let the classical Lagrangian 
for a particle with $n$ degrees of freedom, with generalized coordinates 
$q^k, k=1,\dots n$, be given by
\begin{equation}
 L = g_{ij} \frac{\partial q^i}{\partial t } 
\frac{\partial q^j}{\partial t } + V(q) \EqLabel{Eq1}
\end{equation}
\noindent
The first step in the Lagrangian form of path integral quantization is 
the short time propagator (STP)
\begin{equation}
             (qt \vert q_0 t_0 ) = (2\pi i\hbar )^{-n/2} \bigl(g(q)g(q_0 ) 
             \bigr)^{-1/4} 
             \sqrt{ D} \exp \bigl[ i S(qt,q_0 t_0 )/\hbar \bigr] 
                                                           \EqLabel{Eq2} 
\end{equation}
\noindent
where 
\begin{equation}
               D = \det \left( - \frac{\partial ^2 S }{\partial q^i 
               \partial q^j_0 } \right) \EqLabel{Eq3}
\end{equation}
\noindent
and $ S(qt,q_0 t_0 )$ is the classical action along the classical 
trajectory joining the points $q_1, t_1$ and $q_2,t_2$. It is also the 
generator of canonical transformation corresponding to the time evolution 
. The Lagrangian path integral is obtained by iterating the short time 
propagator $(q_2t_2|q_1t_1)$.

\begin{equation}
      K(qt;q_0 t_0 ) = \lim_{N\rightarrow \infty} \int \prod^{N-1} _{k=1} 
         \rho (q_k )dq_k \prod ^{N-1}_{j=0} (q_{j+1} \epsilon | q_j 0) 
         \EqLabel{Eq4}
\end{equation}
where $\epsilon=t/N$.

The steps that are needed to set up the Hamiltonian path integral quantization 
scheme in arbitrary coordinates and to arrive at a representation for the 
propagator are summarized below.

\vspace{0.5cm}
\noindent
{\bf Short time propagator:} The classical Hamiltonian corresponding to 
\EqRef{Eq1} is given by
\begin{equation}
 H = {g^{ij}p_i p_j\over 2m} + V(q)\,. \EqLabel{Eq5}
\end{equation}
To  define  the  short  time  propagator we employ the generators of time 
evolution  in  terms  of  canonical  variables.  This  definition goes as 
follows.   Let   $t_1,\tau,t_2   $,   with  $t_1  <  \tau  <  t_2  $,  be  
infinitesimally close times.and $q_1,q_2,{\rm and}\, p $ be any values of 
coordinates  and  momenta.  Consider  a  classical  trajectory $\gamma_1$ 
starting  from  $q_1$ at time $t_1$ such that at time $\tau $ its momenta 
are  $p$.Similarly let $\gamma_2$ be the trajectory which has momenta $p$ 
at  time  $\tau$  and  coordinates  $q_2$ at time $t_2$. Next we find the 
generators  $S_{--}(p\tau  t_2,q_1t_1  ) $ and $ S_{++}(q_2t_2,p\tau)$ of 
evolution along the two trajectories $\gamma_1,\gamma_2$ appearing inside 
the respective arguments. These generators are Legendre transforms of the 
classical  action computed along the trajectories $\gamma_1,\gamma_2$ and 
depend  on  the  Hamiltonian  which  we  shall denote by $h(q,p)$.We then 
define the `mixed short time propagators' by 

               \begin{equation} (q_2t_2\vert p\tau) = (2\pi \hbar )^{-n/2} 
               \sqrt{D_{++}} \exp [ i S_{++}(q_2t_2,p\tau)/ 
               \hbar] \EqLabel{Eq6} \end{equation} 

\begin{equation} (p\tau\vert q_1t_1) = (2\pi\hbar )^{-n/2} \sqrt{ D_{--
}}\exp [ i S_{--}(p\tau,q_1t_1)/\hbar ]  \EqLabel{Eq7} 
\end{equation}
where 
\begin{equation} D_{++}= \det \left(\frac{ \partial^2 S_{++}}{ \partial 
q_2^i \partial p_{j}}\right) \EqLabel{(2.7) }\end{equation}
\begin{equation} D_{--} = \det \left( \frac{\partial ^2 S_{--}}{ \partial 
q_1^i \partial p_{j}} \right) \EqLabel{Eq8} \end{equation}

\noindent
and finally the canonical short time propagator is defined by 

\begin{equation} (q_2t_2\Vert q_1t_1) = \frac{1} {\sqrt{\rho (q_1)\rho (q_2)}} 
\int 
d^n p ~ (q_2t_2| p \tau) ~ (p\tau\vert q_1 t_1) \EqLabel{Eq9} 
\end{equation}
\noindent
which propagate the square integrable wave functions $\psi(q)$ with 
measure $\rho(q) d^nq.$

\vspace{0.5cm}
\noindent
{\bf Canonical path integral:} As a next step we define a canonical path 
integral built up from the STP $(q_2t_2\parallel q_1t_1)$ given by \EqRef{Eq9}; 
the resulting path integral denoted by $K[h,\rho](q t,q_0 t_0)$ is defined 
for the Hamiltonian $h(q,p)$ as follows.

\begin{equation}
  K[h,\rho](qt;q_0 t_0 )\stackrel{def}{=} \lim_{N\rightarrow \infty} \int \prod^{N-1} 
  _{k=1} \rho (q_k )dq_k \prod ^{N-1}_{j=0} (q_{j+1} \epsilon \Vert q_j 0) 
  \EqLabel{Eq10} 
\end{equation}
This  definition  is  a  natural  generalization  of  the Lagrangian path 
integral  quantization  scheme  of DeWitt described above. But is seen to 
fail  for the `bench mark' case of free particle in two dimensions in $r, 
\theta$ coordinates.Identifying $h(q,p)$ with the classical free particle 
Hamiltonian  does not lead to the correct Schr\"{o}dinger equation in the 
$r,\theta$ variables for the propagator\cite{AKK1}. 

As  already  mentioned  above,  it possible to modify the above scheme by 
incorporating    the    idea    of    {\it   local   scaling   of   time}   
\cite{HOAKIRA,AKIRA} in such a way that it is not necessary to introduce 
ad  hoc  $O(\hbar^2)$  terms  in  the classical Hamiltonian to obtain the 
correct  Schr\"{o}dinger  equation  for  each  set  of  coordinates.  The  
modified   definition  is  more  general  than  just  getting  the  right  
Schr\"{o}dinger  equation  for  the  free particle. In fact it gives us a 
practical  method for relating the path integrals for different problems, 
thereby  helping  us to tackle many exactly solvable potentials by purely 
path  integral  methods  \cite{AKKPS2}. In the next paragrah we introduce 
this modifed scheme resuting in a Hamiltonian path integral with scaling.
 
\vspace{0.5cm}
\noindent
{\bf  Canonial path integral with local scaling of time }:Let $\alpha(q)$ 
be  a  strictly  positive  function  of  $q$.  It  will be called scaling 
function. Given a hamiltonian $H$, define for any real $E>0$, the pseudo-
Hamiltonian by 
\begin{equation}
H_\alpha^E \stackrel{\rm def}{=} \alpha (H-E) \EqLabel{Eq11}
\end{equation}
The Hamiltonian path integral ${\cal K}$ with scaling $\alpha$ is defined to be 
\begin{eqnarray}
\lefteqn {{\cal K}[H,\rho,\alpha](qt,q_00) }\\ 
&&\equiv \sqrt{\alpha(q)\alpha(q_0)} \int_0^\infty{dE\over 2\pi\hbar} 
\exp[-iEt/\hbar]\int_0^\infty d\sigma K[H^E_\alpha,\rho]
(q\sigma,q_0\sigma_0=0) \EqLabel{Eq12} \\
\end{eqnarray}
The original canonical short time propagator appears in the propagator 
$K$ in the right hand side of the above equation calculated for pseudo-
time $\sigma$ by identifying the function $h$ with the pseudo-Hamiltonian 
$H_\alpha^E.$ It can be shown that for $\alpha=1$ the path integral with 
scaling coincides with the Hamiltonian path integral defined above 
in \EqRef{Eq10}. If we take $H$ as in \EqRef{Eq5} then ${\cal K}$ satisfies the 
following equation for arbitrary $\alpha$ \cite{AKKPS1}. 
\begin{equation}
i\hbar \frac{\partial {\cal K}}{\partial t} =\hat{H}_{\rho,\alpha} {\cal K} \EqLabel{Eq13}
\end{equation}
where
\begin{equation}
\hat{H}_{\rho,\alpha}=- \frac{\hbar^2}{2m}\rho^{-{1\over 2}}\alpha^{-{1\over 
2}} \left(\frac{\partial }{\partial q^i} g^{ij} \alpha 
\frac{\partial}{\partial q^j}\right)\rho^{1\over 2}\alpha^{-{1\over 2}} + 
V \EqLabel{Eq14}
\end{equation}
and has the normalization 
\begin{equation}
\lim_{t\rightarrow t_0} {\cal K}[H,\rho,\alpha](qt,q_0t_0) =
 {1\over\rho(q_0)}\delta^n(q-q_0) \EqLabel{Eq15}
\end{equation}

\noindent
To obtain the correct quantization scheme in arbitrary coordinates one 
needs to select $\alpha=\sqrt{g} ( g=\det (g_{ij}) =1/ \det(g^{ij})$. With 
this choice of the scaling function the Hamiltonian operator in \EqRef{Eq13} 
becomes 
\begin{equation}
\hat{H}_{\sqrt{g},\sqrt{g}} = - \frac{\hbar^2}{2m} {1\over \sqrt{g}}
\left(\frac{\partial }{\partial q^i}g^{ij}\sqrt{g}\frac{\partial 
}{\partial q^j}\right) + V \EqLabel{Eq16}
\end{equation}
which has the invariant Laplace Beltrami operator as the kinetic energy 
part of the Hamiltonian operator. 
It should be noted that the Hamiltonian path integral with scaling is not 
obtained from any short time propagator but from the full finite 
Hamiltonian path integral $K.$

\section{Propagator in plane polar coordinates}
We  shall  illustrate  the  detailed  calculation  of the Schr\"{o}dinger 
equation  for  ${\cal  K}[H,\sqrt{g},\sqrt{g}]$  for  the  case  of  free  
particle in two dimensions in $r\theta$ coordinates. For this problem the 
momenta conjugate to $(r,\theta)$ will be denoted by $(P,p)$ and
\begin{eqnarray}
H={P^2\over2m} + {p^2\over 2mr^2}\\
g^{ij} = \left( \begin{array}{c c}1 & 0 \ \\ 0 & r^{-2} \end{array}\right)\\
\rho=\sqrt{g}=r \EqLabel{Eq17}\\
H^E_{\sqrt{g}} =r(H-E)= {rP^2 \over 2m} + {p^2 \over 2m r} - E r \equiv h
 \EqLabel{Eq18}
\end{eqnarray}
Let 
\begin{equation}
\sigma_0 =0, \sigma_1=\epsilon, \sigma_2=2\epsilon, \dots , \sigma_N = N 
\epsilon = \sigma \EqLabel{Eq19}
\end{equation}
be  the pseudo time grid or slicing for the interval $(0,\sigma)$. The p-
integrations  are  placed  midway  between $\sigma_j$ and $\sigma_{j+1}$, 
i.e.,   at   ${1\over  2}\epsilon,{3\over  2}  \epsilon  ,(j+{1\over  2})  
\epsilon,  ...  $. To first order in $\epsilon,$ if $r_j$ and $\theta_j $ 
are coordinates chosen at $\sigma_j$, $S_{\pm \pm}$ are given by 
\begin{eqnarray}
S_{++}(r_{j+1}\theta_{j+1},P_j,p_j) \approx P_j r_{j+1} + p_j
 \theta_{j+1} - { \epsilon \over 2}h_{j+1} \EqLabel{Eq20}\\
S_{--}(P_j,p_j,r_j\theta_j) \approx -P_j r_j - p_j \theta_j- {\epsilon 
\over 2} h_j \EqLabel{Eq21}\\
h_{j+1}=\frac{r_{j+1}P_j^2}{2m} + \frac{p_j^2}{2mr_{j+1}} -Er_{j+1} \EqLabel{Eq22}\\
h_j=\frac{r_jP_j^2}{2m} + \frac{p_j^2}{2mr_j} -Er_j \EqLabel{Eq23}
\end{eqnarray}
The factors $D_{\pm\pm}$ do not trouble us because these are equal to 
$1+O(\epsilon^2).$ The short time 
propagator is 
\begin{equation}
(r_{j+1}\theta_{j+1}\parallel r_j\theta_j) = 
{1\over(2\pi\hbar)^2}\int dP_j \int dp_j {1\over\sqrt{r_{j+1}r_j}} 
\exp[iS_j / \hbar] \EqLabel{Eq24}\\
\end{equation}
\begin{equation}
S_j=P_j(r_{j+1}-r_j) +p_j(\theta_{j+1}-\theta_j) -{\epsilon\over 
2}(h_{j+1}-h_j )\EqLabel{Eq25}
\end{equation}
and 
\begin{eqnarray}
\lefteqn{ K[h,\sqrt{g}](r,\theta,\sigma,r_0,\theta_0\sigma_0=0)}\\ 
&&=\lim_{N\rightarrow\infty} \int \big(\prod_{k=1}^{N-1} r_k dr_k 
d\theta_k \big)
(r\theta\epsilon\parallel r_{N-1}\theta_{N-1}0) ....
(r_1\theta_1\epsilon\parallel r_0\theta_0 0) \nonumber \\ 
&&=\lim_{N\rightarrow\infty} \int \int d(N-1) ...\int d(1) dP_0 dp_0 
\exp\big[ i\sum_{j=0}^{N-1} S_j/\hbar\big] \EqLabel{Eq26}
\end{eqnarray}
where
\begin{eqnarray}
r_N=r ; \qquad \theta_N=\theta ; \qquad\qquad\EqLabel{Eq27}\\
d(j) =r_j dr_j d\theta_jdP_jdp_j/(2\pi\hbar^2)\EqLabel{Eq28}
\end{eqnarray}
\noindent
Since we are interested only in deriving the Schr\"{o}dinger equation, for our 
purpose it is sufficient to compute the propagator ${\cal K}[H,\sqrt{g},\sqrt{g}] 
$ for small $t$ only. The expression for ${\cal K}$ for the short times becomes 
\begin{eqnarray}
\lefteqn {{\cal K}[H,\sqrt{g},\sqrt{g}](r\theta t,r_0\theta_0 0)} \nonumber \\
&& =\lim_{N\rightarrow\infty}\sqrt{rr_0} \int_0^\infty {dE\over 
2\pi\hbar} \exp(-iEt/\hbar) \nonumber \int_0^{\infty} d\sigma\\ && 
\hspace{2cm} \int d(N-1)...d(1) \int dP_0 dp_0 \exp\LARGE\{ 
i\sum_{j=0}^{N-1}S_j/\hbar\LARGE\} 
\EqLabel{Eq29}
\end{eqnarray}
Note that each $h_j$ in $S_j$ carries a term $-Er_j$ Thus $E$ integration 
can be done immediately to give the delta function
\begin{equation}
\delta\big[\,\epsilon ({\mit {1\over 2}}(r+r_0) +r_1+...+r_{N-1}) -t\, \big]
 \EqLabel{Eq30}
\end{equation}
\noindent
which allows us to do the $\sigma$ integration $(\epsilon=\sigma/N)$. The 
net result is that we get an overall factor 
\begin{equation}
{2N\over F} \equiv \frac{2N}{r_0+2(r_1+...+r_{N-1}) + r } \EqLabel{Eq31}
\end{equation}
and $\epsilon\,\sigma/N$ is replaced by $2t/F.$ Thus 
\begin{equation}
{\cal K}_t=\int dP_0 dp_0 \int d(N-1) .... d(1)\left({2N\over 
F}\right) \nonumber\\ \exp\left[ i \sum S_j / \hbar \right]\EqLabel{Eq32}
\end{equation}
with
\begin{equation}
S_j = P_j(r_{j+1}-r_j) + p_j(\theta_{j+1} 
-\theta_j) - {t\over F} \left( \frac{P_j^2}{2m}(r_{j+1}+r_j) +
\frac{p^2_j}{2m}({1\over r_{j+1}}+{1\over r_j}) \right)\EqLabel{Eq33}
\end{equation}
Here on we write ${\cal K}_t$ for ${\cal K}[H,\sqrt{g},\sqrt{g}] $, the arguments being 
suppressed.We shall also omit the explicit mention of limit $N 
\rightarrow \infty$, as this limit will be taken in the end only.

\vspace{0.5cm}
\noindent
{\bf Propagator} ${\cal K}$ {\bf for short times:} We are interested in the 
Schr\"{o}dinger equation. For this purpose it is sufficient to take $t$ 
infinitesimally small. For $t$ actually equal to zero the propagator ${\cal K}_t$ 
becomes 
\begin{equation}
{\cal K}_{t=0} = \int dP_0 dp_0 \int d(N-1)..\int d(1) 
  {2N\over F}\exp\left(i \sum_j S_j/\hbar \right)\EqLabel{Eq34}
\end{equation}
The $p-$ integrations can be carried out and one gets
\begin{eqnarray}
\lefteqn {{\cal K}_{t=0}}
&=& \int\left[ {\small \prod}_{j=1}^{N-1} dr_j d\theta_j\right] {2N\over F} 
\delta(r-r_{N-1}) \delta(\theta-\theta_{N-1}) .... \delta(r_1-r_0) 
\delta(\theta_1-\theta_0) \nonumber \cr
&=&{1\over r} \delta(r-r_0) \delta(\theta-\theta_0) \EqLabel{Eq35}
\end{eqnarray}
\noindent
as $F \rightarrow 2Nr $ when all $r_1=r_2=...=r_{N-1}=r$. For finite but 
small t, the exponential is expanded to first order in t;
\begin{eqnarray}
\lefteqn {{\cal K}_t-{\cal K}_0 } \nonumber \\&& \equiv
\left({-it\over \hbar}\right) \sum_k \int dP_0 dp_0 \int d(N-1)...d(1) 
{2N\over F^2} \times \nonumber \\&&
\hspace{2 cm}\left( \frac{P_k^2}{2m}(r_{k+1}+r_k) +
\frac{p^2_k}{2m}({1\over r_{k+1}}+{1\over r_k}) \right) \times \nonumber \\&&
\hspace{2cm} \exp\left[ {i\over \hbar} \sum_{j=0}^{N-1}\left\{ P_j(r_{j+1}
 -r_j) + p_j (\theta_{j+1} -\theta_j) \right\}\right] \nonumber \\&&
\equiv \left({-it\over \hbar}\right) \sum_k X_k \EqLabel{Eq36}
\end{eqnarray}
where 
\begin{eqnarray}
\lefteqn{ X_k = } && \int\! dP_0 dp_0 \int \!d(N-1)...d(1) {2N\over F^2} 
\left( \frac{P_k^2}{2m}(r_{k+1}+r_k) +
\frac{p^2_k}{2m}({1\over r_{k+1}}+{1\over r_k}) \right) \nonumber \\ 
&& \exp\left[ {i\over \hbar} \sum_{j=0}^{N-1}\left\{ P_j(r_{j+1} -r_j)
+ p_j (\theta_{j+1} -\theta_j) \right\}\right] \EqLabel{Eq37}
\end{eqnarray}
\vspace{0.5cm}
\noindent
{\bf Computation of $X_k$ }\,: All the momenta integrations in $X_k$, 
except over the $ k$-th momenta, can be done giving $2(N-1)\,\, \delta $ 
functions.
\begin{eqnarray}
\delta(r-r_{N-1})...\delta(r_{k+2}-r_{k+1})\delta(r_{k}-r_{k-1})...
\delta(r_1-r_0) \hfill\EqLabel{Eq38} \\
\delta(\theta -\theta_{N-1})... \delta(\theta_{k+2} -\theta_{k+1})
\delta(\theta{k}-\theta{k-1}) ...\delta(\theta_1 -\theta_0) \hfill\EqLabel{Eq39}
\end{eqnarray}
\noindent
This permits \underline {all} $r,\theta$ integrations to be done resulting 
in the replacements

\begin{equation}
r=r_{N-1}=...=r_{k+1}; \,\,\,\, r_0=r_1=...=r_k \EqLabel{Eq40}
\end{equation}
\noindent
So
\begin{equation}
X_k = \int dP dp\, \frac{2N}{F^2_k} \left[ \frac{P^2}{2m}(r+r_0)+
\frac{p^2}{2m}({1\over r}+{1\over r_0}) \right] 
\exp\left[{i\over\hbar}\left( P(r-r_0)+p(\theta-\theta_0) \right) \right]
 \EqLabel{Eq41}
\end{equation}
\noindent
where we have renamed $P_k,p_k$ as $P$ and $p$ respectively. The $k-$ dependence of 
$X_k$ resides only in $F_k$ which is obtained from $F$ 
after replacements as in \EqRef{Eq38}
\begin{eqnarray}
F_k&=& r+2((N-k-1)r+kr_0)+r_0 \hfill\nonumber \\
   &=&(2N-1-2k)r+(2k+1)r_0 \hfill \EqLabel{Eq42}
\end{eqnarray}
Writing
\begin{equation}
F_k^{-2} = \int_0^\infty d\beta \beta \exp(-\beta F_k) \EqLabel{Eq43}
\end{equation}
\begin{eqnarray}
\lefteqn{X_k} && =
 2N \int dP dp \left[ \frac{P^2}{2m}(r+r_0) + \frac{p^2}{2m}\left 
({1\over r}+{1\over r_0}\right)\right]
\exp\left[{i\over\hbar}\left( P(r-r_0)+p(\theta-\theta_0) \right) \right]
 \times \nonumber \\ &&
\hspace{1cm} \int_0^\infty d\beta\, \beta \exp[-\beta r (2N-2k-1) - \beta 
r_0 (2k+1) ] \nonumber 
\\&&= 2N \int_0^\infty d\beta\,\beta \exp[-\beta r (2N-2k-1) - \beta r_0 
(2k+1)]\times \nonumber 
\\ &&\hspace{1cm} -\big(\frac{\hbar^2}{2m}\big)[\,(r+r_0)\delta^{\prime\prime}(r-
r_0) 
 \delta(\theta-\theta_0)+ \nonumber 
\\ && \hspace{3cm} ({1\over r}+{1\over r_0})\delta(r-r_0)
 \delta^{\prime\prime}(\theta-\theta_0)\, ]\EqLabel{Eq44}
\end{eqnarray}
where in the last step $P$ and $p$ integrations have been computed. 

\vspace{0.5cm}
\noindent
{\bf Schr\"{o}dinger equation :}The functions propagated in time by {\cal K} 
\begin{equation}
\psi(r, \theta,t) = \int {\cal K}[H,\sqrt{g},\sqrt{g}](r\theta t,r\theta 0) 
\psi(r_0, \theta_0) r_0 dr_0 d\theta_0\EqLabel{Eq45}
\end{equation}
for small $t$, satisfy the equation

\begin{eqnarray}
i\hbar \frac{\partial \psi}{\partial t} = i \hbar \int r_0 dr_0 d\theta_0 
\left( \frac{{\cal K}_t-{\cal K}_0}{t}\right) \psi(r_0 \theta_0) \\
= \sum_k \int r_0 dr_0 d\theta_0 X_k(r\theta,r_0\theta_0) \psi(r_0 
\theta_0) \EqLabel{Eq46}
\end{eqnarray}
\noindent The integrations over $r_0,\theta_0$ are trivial in view of the 
$\delta -$ functions. For any k,
\begin{eqnarray}
\lefteqn { \int r_0 dr_0 d\theta_0 X_k(r\theta,r_0\theta_0) 
\psi(r_0\theta_0)}\\ &&
= -{\hbar^2\over 2m} (2N) \int d\beta e^{-2\beta r N } 
{\partial^2 \psi\over \partial \theta^2}\\ &&
\left. - {\hbar^2\over 2m} (2N) \int d\beta 
e^{-\beta r (2N-2k-1)} \frac{\partial^2}{\partial r_0^2}\left((r+r_0) 
e^{-\beta r_0(2k+1)} r_0\psi \right) \right|_{r=r_0} \EqLabel{Eq47}
\end{eqnarray} 
The first term on summing over $k$ becomes
\begin{equation}
\sum_{k=0}^{N-1} 2N \int d\beta\beta \exp(-2\beta r N)\, 2\, \frac
 {\partial^2 \psi }{\partial \theta^2} 
= {1\over r^2}\frac{\partial^2\psi }{\partial \theta^2}\EqLabel{Eq48}
\end{equation}

For the second term, we note that 
\begin{eqnarray}
\frac{\partial^2}{\partial r_0^2}
{ \LARGE [ } {\big( }r_0 r +r_0^2 {\big )} e^{-\beta r_0(2k+1)}\psi {\LARGE ]} 
= e^{-\beta r_0(2k+1)} {\LARGE \{ } -6\beta r(2k+1) \psi \hspace{2cm}\nonumber \\
\ \ \ + 2\beta^2r^2(2k+1)^2 \psi +2 \psi 
+(-4 \beta r^2 (2k+1)+ 6r)\frac{\partial \psi}{\partial r} + 2 r^2 
\frac{\partial^2 \psi}{\partial r^2} {\LARGE \}} \EqLabel{Eq49}
\end{eqnarray}

Therefore, on using\\
\begin{eqnarray}
\sum_{k=0}^{N-1} (2k+1)&=&N^2 \hspace{6cm} \EqLabel{Eq50}\\
\sum_{k=0}^{N-1} (2k+1)^2&=&{1\over 3}N(2N-1)(2N+1)\approx {4\over 3}N^3 \hspace{1cm} 
 \EqLabel{Eq51}\\
\int_0^\infty d\beta \beta^n\exp(-2\beta r N) &=& \frac{n!}{(2 rN)^{n+1}} \hspace{3cm},
\EqLabel{Eq52} \\
\end{eqnarray}
the expression 

$$\sum_k 2N \int_0^\infty d\beta \beta e^{-2\beta r N}{ \LARGE [}
\left(-6\beta r(2k+1) + 2\beta^2r^2(2k+1)^2 +2\right) \psi $$ 
$$ + \left(-4 \beta r^2 (2k+1)+ 6r\right)\frac{\partial \psi}{\partial r}
+ 2 r^2 \frac{\partial^2 \psi}{\partial r^2}{\LARGE ]}$$
becomes
\begin{equation}
{1\over r}{\partial \psi \over \partial r} + \frac{\partial^2 
\psi}{\partial r^2}\EqLabel{Eq53}
\end{equation}
Thus we get the desired Schr\"{o}dinger equation 
\begin{equation}
i \hbar \frac{\partial \psi}{\partial t} = - \frac{\hbar^2}{2m}
\left(\frac{\partial^2\psi}{\partial r^2} + {1\over r} \frac {\partial 
\psi}{\partial r} + \frac{\partial^2 \psi}{\partial \theta^2} 
\right)\EqLabel{Eq54}
\end{equation}
which  shows  that  use  of  ${\cal K}[H,\sqrt{g},\sqrt{g}]$ leads to the 
correct propagator for these coordintes. The expression for ${\cal K}$ is 
not  yet  the  final  proppagator,  one  has  to take into account of the 
boundary condistions at $r=0$ and for $\theta=0,2\pi$. This can be easily 
done and the final answer for the propagator is 

\begin{eqnarray}
\lefteqn{{\cal K}(r,\theta t; r_0,\theta_0, 0,) = } && \nonumber
\\ && 
\sum_{m=-\infty}^{\infty}\left[ {\cal K}(r,\theta+2\pi m, t, r_0,\theta_0, 0,) +
{\cal K}(-r,\theta+ 2\pi(2m+1),t, r_0,\theta_0, 0,)\right]\nonumber\\
\end{eqnarray}

\end{document}